\documentclass[prd,aps,amsmath,amssymb,nofootinbib,twocolumn,preprintnumbers]{revtex4}

\usepackage{graphicx}
\usepackage{dcolumn}
\usepackage{bm}
\usepackage{amsmath}
\usepackage{amsthm}
\usepackage{amsfonts}
\usepackage{euscript,bbm}
\usepackage{ifthen}
\usepackage{psfrag}
\usepackage{slashed}
\usepackage{hyperref}
\usepackage{color}
\usepackage{url}

\def\ls{\mathrel{\lower4pt\vbox{\lineskip=0pt\baselineskip=0pt
           \hbox{$<$}\hbox{$\sim$}}}}
\def\gs{\mathrel{\lower4pt\vbox{\lineskip=0pt\baselineskip=0pt
           \hbox{$>$}\hbox{$\sim$}}}}
\def\drawbox#1#2{\hrule height#2pt
\hbox{\vrule width#2pt height#1pt \kern#1pt
              \vrule width#2pt}
              \hrule height#2pt}

\def\Asym#1#2{\vcenter{\vbox{\drawbox{#1}{#2}
              \kern-#2pt       
              \drawbox{#1}{#2}}}}

\newcommand{\be}{\begin{equation}}
\newcommand{\ee}{\end{equation}}
\newcommand{\bea}{\begin{eqnarray}}
\newcommand{\eea}{\end{eqnarray}}

\newcommand{\neu}[1]{\ensuremath{\tilde{\chi}_{#1}^0}}

\newcommand{\st}{\ensuremath{\tilde{t}}}

\newcommand{\gsim}{\lower.7ex\hbox{$\;\stackrel{\textstyle>}{\sim}\;$}}
\newcommand{\lsim}{\lower.7ex\hbox{$\;\stackrel{\textstyle<}{\sim}\;$}}

\newcommand{\met}{{E\!\!\!\!/_{\rm T}}}

\newcommand{\pythia}{{\tt Pythia v6.426}}

\newcommand{\pgs}{{\tt PGS4}}

\newcommand{\madgraph}{{\tt MadGraph5 v2.2.2}}

\newcommand{\ben}{\begin{enumerate}}
\newcommand{\een}{\end{enumerate}}
\newcommand{\bei}{\begin{itemize}}
\newcommand{\eei}{\end{itemize}}

\begin{document}

MI-TH-1518

\title{Dilepton Mass Endpoint in the NMSSM}

\author{Bhaskar Dutta$^{1}$}
\author{Yu Gao$^{1}$}
\author{Tathagata Ghosh$^{1}$}
\author{Teruki Kamon$^{1,2}$}
\author{Nikolay Kolev$^{3}$}
\affiliation{
$^{1}$~Department of Physics and Astronomy, Mitchell Institute for Fundamental Physics and Astronomy, Texas A{\&}M University, College Station, TX 77843-4242, USA\\
$^{2}$~Department of Physics, Kyungpook National University, Daegu 702-701, South Korea\\
$^{3}$~Department of Physics, University of Regina, SK, S4S 0A2, Canada
}

\begin{abstract}
NMSSM scenarios are investigated to explain an excess in the opposite-sign dilepton mass distribution in events with dilepton, jets and missing transverse energy reported by the CMS experiment. We show that the NMSSM scenarios can possess unique features  to explain this excess, and can be distinguished from the MSSM scenarios in the ongoing LHC runs as well as direct detection experiments. 
\end{abstract}

\maketitle

\section{Introduction}

Recently, CMS reported of an excess of lepton pairs~\cite{bib:CMSlepEdge} with energy below the $Z$ mass, in a final state of $l^+l^-jj+\met$ where $\met$ denotes the missing transverse energy. Beyond the the Standard Model (BSM) theoretical attempts have been made to account for such an excess in the context of the Minimal Supersymmetric Standard Model (MSSM)~\cite{bib:CMSlepEdge,Huang:2014oza,Grothaus:2015yha} using cascade decays from the sbottom pair production at the LHC, a leptoquark scenario~\cite{Allanach:2015ria} and superstring inspired models~\cite{Dhuria:2015hta}, where the cascade decay of  new particle states give rise to a pair of leptons, two associated jets, and new invisible particles that escape the detector as $\met$.   

Two same-flavor, opposite-sign leptons can be produced from a cascade decay that has  particles decay into leptons in the intermediate state during the process. Such lepton partners and heavier states, which give rise to the cascades, are readily available in supersymmetric models.  The sbottom pair production and its cascade decays into the next to lightest neutralino which subsequently decays into two leptons and the lightest neutralino via an intermediate slepton state is  a very interesting option to explain the excess. Two $b$ jets are also produced along with $l^+l^-+\met$ in the final states. The existence of two $b$ jets in the signal provides an interesting prediction arising from this scenario which will be checked in Run II. Prior to this new result, $l^+l^-jj+\met$ was considered as a possible final state from the stop decay~\cite{dutta}. 
Since we have $\st \rightarrow t+\neu{2}$, we expect lepton(s) from top decay in addition to $\neu{2}\rightarrow l^+ l^- \neu{1}$. This doesn’t support the CMS edge paper. In this paper we focus on a well motivated Next-to Minimal Supersymmetric Standard Model~\cite{bib:nmssm} (NMSSM), which introduces an additional singlet superfield into the MSSM.  The observed Higgs mass at the LHC~\cite{bib:LHCHiggs} can be accommodated  naturally if the coupling $\lambda$ between the singlet and the supersymmetric Higgs fields is large and this new term also provides a solution to  the $\mu$ problem of the MSSM (see~\cite{NMSSM:report} and references therein). Recently, signals of such a scenario at the LHC were investigated and possible ways to distinguish from the MSSM were also discussed~\cite{Dutta:2014hma}. 

The singlet superfield in the NMSSM gives rise to a new neutralino (singlino) to the gaugino sector, besides other modification on the MSSM particle spectrum. This paper investigates the effect of the singlino that creates more freedom to realize the aforementioned cascade decays of sbottom, which alleviates a relatively tight requirement on the MSSM gaugino mixings. Further, since the neutralino sector is modified the NMSSM explanation will be associated with distinguishable predictions.  The cascade process to explain the endpoint in the context of the NMSSM is discussed in Section~\ref{sect:cascade}. A collider analysis is carried out in Section~\ref{sect:collider} that examines the NMSSM's explanation of the CMS dilepton excess. We discuss our results and  conclude in Section~\ref{sect:discussions}.

\begin{figure}
\includegraphics[scale=0.6]{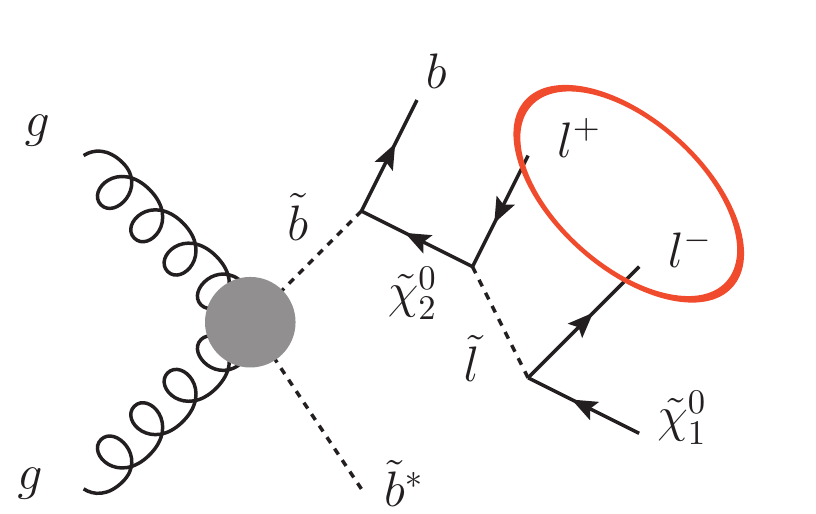}
\caption{Pair production of sbottoms and one of its cascade decay chains can give rise to a signal of lepton pairs with energy below the neutralino mass difference. The other sbottom in the diagram presumably decays directly to $\tilde{\chi}^0_1 b$.}
\label{fig:feynman}
\end{figure}

\section{Squark-Neutralino cascades}
\label{sect:cascade}
The CMS excess in the relatively low energy `endpoint' in opposite sign dileptons can rise from the cascade decays of one or more next-to-lightest supersymmetric particles (NLSPs) in the supersymmetric neutralino sector, where the second lightest neutralino can decay via a two-step process,
\be 
\tilde{\chi}^0_2\rightarrow l^+ \tilde{l}^-,~\tilde{l}^-\rightarrow l^-\tilde{\chi}^0_1,
\ee
that yields a pair of leptons of the same flavor, whose total energy distribution is limited by the mass difference $M_{\tilde{\chi}^0_2}-M_{\tilde{\chi}^0_1}$ ($\tilde{l}$ denotes both both left and right sleptons of the first two lepton flavors.). Thus a neutralino spectrum with $M_{\tilde{\chi}^0_2}\sim M_{\tilde{\chi}^0_1} +70$ GeV can lead to such a dilepton excess if the slepton masses lie between the two neutralino masses and if the second neutralino can be produced at adequate rates.

The QCD-dominated production of sbottom\footnote{The dilepton mass endpoint can also be produced from the decay of a light stop, but when stop is pair produced we also expect to see multilepton final state and the non-observation of more than two leptons  along with 2 jet and $\met$ makes this scenario not preferred.} can lead to the correct pair-production rates for the CMS experiment.
 A mostly right-handed sbottom (denoted by $\tilde{b}_1$ here) can be a perfect candidate~\cite{bib:CMSlepEdge} that decays into the required neutralinos. The  Feynman diagram for the production process is illustrated in Fig.~\ref{fig:feynman}. The sbottom is preferably right-handed as it does not require a left-handed stop to be equally light in mass.
Moreover, the CMS non-observation of more than two leptons in this channel puts a constraint on the decay of such a sbottom, i.e., the branching fraction, BF$(\tilde{b}_1\rightarrow\tilde{\chi}^0_1 b)$, should dominate over BF$(\tilde{b}_1\rightarrow\tilde{\chi}^0_2 b)$ unless the latter produces  many lepton pairs in a final state from the pair of sbottoms. This requirement, however, may force certain relations among the neutralino mixings.

In the MSSM, a sbottom can decay into a $b$ quark and a neutralino via either the U(1)$_Y$ gauge coupling or the Yukawa coupling. The U(1)$_Y$ gauge coupling depends on the Bino ($\tilde{B}$) component of the daughter neutralino, while the Yukawa coupling depends on the down-type Higgsino ($\tilde{H}_d$) component and the size of the coupling $y_d$. Since the MSSM's LSP needs to be dominantly Bino to avoid increasingly severe experimental constraints (from indirect~\cite{wh-ID,Ajaib} and direct~\cite{higgsino-DD} detections). However, the Bino content of the LSP would dictate the sbottom decay (by a few orders of magnitude) and leaves with  small $\tilde{B}$ for $\tilde{\chi}^0_2$ for the $\tilde{b}_1\rightarrow\tilde{\chi}^0_2 b$ decay when $M_{\tilde{\chi}^0_2}-M_{\tilde{\chi}^0_1}\approx 70$ GeV. The BF to $\tilde{b}_1\rightarrow\tilde{\chi}^0_2 b$ can be increased when $\tan\beta$ is large which causes larger Yukawa coupling contribution~\cite{Huang:2014oza}.

In the NMSSM, however, the option of a singlino LSP opens up alternative gaugino mixing scenarios to realize sbottom decays into both $\tilde{\chi}^0_1$ and $\tilde{\chi}^0_2$ in a suitable way, which differ from the MSSM. Consequently, the collider signals arising from the direct production of gauginos will be helpful to discern MSSM from NMSSM, 

(A) We can satisfy the CMS excess for  mostly  singlino LSP and a heavy Bino. The components of $\tilde B$ in both $\tilde{\chi}^0_1$,$\tilde{\chi}^0_2$ are small but comparable in magnitude for heavier $\tilde B$. This allows a small $\tan\beta$ scheme where both $\tilde{b}_1\rightarrow\tilde{\chi}^0_1 b,\tilde{\chi}^0_2 b$ decays occur via the U(1)$_Y$ coupling, while the down-type Yukawa contribution to the decays is small.

The NLSP can be wino and/or Higgsino. A wino-NLSP case is shown in Table~\ref{tab:benchmarks}. In comparison, when the NLSP is mainly Higgsino, $\tilde{\chi}^0_3$ would be relatively light and the sbottom may decay into two NLSPs. However, the Higgsino-NLSP case tends to allow the LSP to have a larger Higgsino mixing which faces constraint from the direct detection result unless we chose the correct sign of gaugino, Higgsino mass parameters to cancel the Higgs contribution in the direct detection amplitude~\cite{dutta3}.

\begin{table}[h]
\begin{tabular}{c|cccccc|ccc|cc}
\hline \hline
\ \ \ \ \ & $M_1$ & $M_2$ & $\tan\beta$ & $\lambda$ & $\kappa$ &$\mu_{\text{eff}}$ &$M_{\tilde{\chi}^0_1}$ &$M_{\tilde{\chi}^0_2}$ &$M_{\tilde{b}_1}$ &$M_{\tilde{\chi}^+_1}$ &$M_{\tilde{l}}$ \\
\hline
A & 500 & 315 & 3.1 & 0.7 & 0.143 & 600& 250 & 320 & 373 & 315 & 285 \\
B & 248 & 800 & 15& 0.5& 0.265& 330& 230& 300&357 &333 &265\\
C & 310 & 800 & 10& 0.6& 0.14& 500& 229& 305&357 &500 &850\\
\hline \hline
\end{tabular}
\caption{The NMSSM benchmark points that yield same-flavor opposite-sign lepton pairs in sbottom cascade decays. The mass spectrum is evaluated using NMSSMTools~\cite{bib:nmssmtools} and its values are given in GeV. $\tilde{l}$ denotes both both left and right sleptons of the first two lepton flavors.}
\label{tab:benchmarks}
\end{table}

(B) We can satisfy the CMS excess for  mostly Bino type $\tilde{\chi}^0_1$, where $\tilde{\chi}^0_2$ is mostly singlino and  $\tilde{\chi}^0_{3,4}$  consist of mostly Higgsinos and Wino. Just like the MSSM, a large $\tan\beta$ is required to boost the decay via down-type Yukawa coupling to the $\tilde{H}_d$ component in $\tilde{\chi}^0_2$, which is closer in mass to the Higgsinos in comparison to the much lighter $\tilde{\chi}^0_1$. The major difference of this scenario from the MSSM is that the lightest chargino $\tilde{\chi}^+_1$ mass is close to $\tilde{\chi}^0_3$ rather than $\tilde{\chi}^0_2$, and consequently can be heavier than that of the MSSM. This  allows a wider mass range of the sbottom after satisfying $M_{\tilde{b}_1}-M_{\tilde{\chi}^+_1}<M_{t}$  so that $\tilde{b}_1$ does not decay into top quarks.

(C) We can also satisfy the CMS excess for a mostly singlino type  $\tilde{\chi}^0_1$ where  $\tilde{\chi}^0_2$ is mostly Bino. 
Since  $\tilde{\chi}^0_2$ is mostly Bino,  BF$(\tilde{b}_1\rightarrow\tilde{\chi}^0_2 b)$ is large and  yields a large number of final state leptons via sleptons situated in between the two neutralinos which may not be a suitable option. A virtual slepton mediated three-body decay \{$\tilde{\chi}^0_2\rightarrow \tilde{l}^*l^+,\tilde{l}^*\rightarrow l^-\tilde{\chi}^0_1$\}, however,  can  give the correct $2l+\met$ rate. This parameter space is represented by point C which shows that the slepton masses are  much higher than the $\tilde{\chi}^0_{1,2}$ mass range, e.g., the slepton masses are almost at TeV scale for point C. It is interesting to note that even if we imagine  a scenario where the slepton masses are very close to either $\tilde{\chi}^0_2$ or $\tilde{\chi}^0_1$, and only one of the leptons from each sbottom cascade is visible, the invariant mass of the two leptons from different cascades can easily be more than 70 GeV and contradicts with observation.

\section{Collider Signal}
\label{sect:collider}
Here we discuss the dilepton yield from the NMSSM benchmark points at the 8 TeV LHC. At each point, a mass gap is kept at 70 GeV between the two lightest neutralinos that limits the energy of the lepton pair. $\tilde{\chi}^0_3$ is above $\tilde{b}_1$ and all sfermions other than $\tilde{l}$ and $\tilde{b}_1$ have multi-TeV masses and decouple from our study\footnote{We remain agnostic about the exact mass of other heavy squarks and ignore the small change in the NLO $\tilde{b}$ production cross-section due to variation of them.}. The lightest chargino $\tilde{\chi}^+_1$ mass is  not lighter than $M_{\tilde{b}_1}-M_t$, which can be used as one aspect that the NMSSM spectrum differs from the MSSM's. The $\tilde{b}_1$ is mostly right-handed and its decay BFs are listed in Table~\ref{tab:BFs}, together with the BF of $\neu{2}$'s decay to $\tilde{l}$ and the lepton reconstruction efficiency. The BF($\tilde{l} \rightarrow l \neu{1}$) decay is $100\%$ for all of our benchmark points, since in all these cases the LSP is either a singlino or bino.

\begin{table}[h]
\begin{tabular}{c|cc|c|c}
\hline \hline
\ \ \ \ \ \ \ & BF$_{\tilde{b}_1\rightarrow\tilde{\chi}^0_1 b}$ & BF$_{\tilde{b}_1\rightarrow\tilde{\chi}^0_2 b}$ & 
BF$_{\tilde{\chi}^0_2 \rightarrow l \tilde{l}}$ & $\epsilon_{ll}$ \\
\hline
A &69\% &31\% &25\% &38\% \\
B &90\% &10\% &57\% &38\% \\
C &10\%  &89\% &11\% &26\% \\
\hline \hline
\end{tabular}
\caption{$\tilde{b}_1$ decay branching fractions and the dilepton selection efficiencies for the benchmark points. $\epsilon_{ll}$ is given in Eq.~\ref{eq:est}. 
BF$_{\tilde{\chi}^0_2 \rightarrow l \tilde{l}}$ for point C in column 4 should read as BF$_{\neu{2} \rightarrow ll\neu{1}}$.}
\label{tab:BFs}
\end{table}

We generate inclusive sbottom pair production events with 0-2 associated jets at 8~TeV in \madgraph~\cite{bib:Madgraph5} with {\tt CTEQ6.6}~\cite{CTEQ6.6} parton distribution functions. \pythia~\cite{bib:Pythia6} is used for showering and hadronization and \pgs~\cite{bib:PGS4} is used for detector simulation in which the electron and muon detection efficiencies are assumed to be  92\% and 98\%, respectively. To avoid double-counting of jets the MLM jet matching scheme~\cite{MLM} is implemented.

The LO production cross section $\sigma_{\text{LO}} \approx 500$~fb for sbottom with approximately 360 GeV mass. This is scaled up by a K-factor of 1.7 to the NLO value that is obtained from the package  {\tt Prospino}~\cite{bib:ProspinoCode}. The dilepton signal rate is then,
\be 
(20~\text{fb}^{-1})\cdot \sigma_{\text{NLO}} \cdot  A_{\rm{eff}},
\ee
where $A_{\rm{eff}}$ denotes the total event selection acceptance. 
The selection criteria for the jet and lepton objects are as follows: $p_T$(jet)$>40$ GeV with $|\eta| < 3$, $p_T$(lepton)$>20$ GeV with $|\eta| < 2.4$, excluding $1.4 < |\eta| <1.6$; the central region is defined as  $|\eta| <1.4$.

\begin{table}[h]
\begin{tabular}{c|c|c|c}
\hline \hline
Event selection& \multicolumn{3}{c}{Relative Efficiency}\\
\cline{2-4}
& A & B & C\\
$\sigma_{\rm{NLO}}$ (fb)  & 660 & 854 & 854 \\\hline
$N_{j} \geq 2 (3) \,+\, \met> 150 (100) $ GeV & 32\%&37\%&25\%\\
Two isolated OSSF leptons & 3.2\%&2.3\%&3.2\%\\
Dileptons in the central region & 85\%&85\%&84\%\\
\hline
Overall acceptance & 0.85\%&0.72\%&0.66\%\\\hline
Number of events at 20 fb$^{-1}$&112&124&112\\
\hline \hline

\end{tabular}
\caption{The selection efficiencies for the $l^+l^-jj+\met$ signal at benchmark points A, B and C.}
\label{tab:cut}
\end{table}

The selection efficiency flows for points A, B and C are shown in Table~\ref{tab:cut}. Overall acceptances are similar for A and B. The overall acceptance for point C is small due to the fact. The first cut is the ``OR" cut from the CMS analysis~\cite{bib:CMSlepEdge}: it selects events with at least 2 jets and $\met>$150~GeV, or at least 3 jets and $\met>$100~GeV. We find that point C shows lower efficiency at this stage compared to points A and B. This is due to the fact that $\tilde{b}_1$ has a larger branching ratio into $\tilde{b}_1\rightarrow b+\tilde\chi^0_1$ for points A and B where $\tilde\chi^0_1$ is mostly Bino compared to the point C where $\tilde\chi^0_1$ is mostly singlino. A large mass gap ($\sim$ 120 GeV) between $\tilde{b}_1$ and $\tilde\chi^0_1$ causes the existence more higher $p_T$ jets for points A and B. On the other hand, for point C, $\tilde{b}_1$ has larger branching ratio into $b+\tilde\chi^0_2$ and a smaller mass gap ($\sim 50$ GeV) between $\tilde{b}_1$ and $\tilde\chi^0_2$ produces lower $p_T$ jet for the point C which causes lower efficiency after we require $p_T$(jet)$>$40 GeV. The fraction of isolated leptons is a combination of the $\tilde{b},\tilde{\chi}^0_2$ decay branching fractions and dilepton selection efficiency $\epsilon_{ll}$,

\be
\begin{array}{cc}
2\cdot x (1-x) ~\epsilon_{ll}, \\
 x \equiv \text{BF}(\tilde{b}_1\rightarrow\tilde{\chi}^0_2 b)~\text{BF}(\tilde{\chi}^0_2 \rightarrow l \tilde{l}). 
\end{array}
\label{eq:est}
\ee
$\epsilon_{ll}$ is a fraction of events with at least two reconstructed OSSF leptons passing the lepton selection criteria out of the events with two OSSF leptons from the $\tilde{\chi}^0_2$ decays.
In principle, $\epsilon_{ll}$ can vary with the lepton energy that is determined by the mass difference between the sleptons and neutralinos.
At points A and B, when $M_{\tilde{l}}\sim (M_{\tilde{\chi}^0_1}+M_{\tilde{\chi}^0_2})/2$ with both leptons at $E_l=35$~GeV, we get maximal acceptance of $\epsilon_{ll}\approx 38\%$. 
At point C the leptons arise from a three body decay and the energy partition becomes uneven, and the acceptance can suffer if one of the leptons is too soft. 
The lepton pairs are further required to be in the central region with pseudorapidity $|\eta| < 1.4$, and finally the overall acceptance gives the accumulated selection efficiency.

\begin{figure}[t]
\includegraphics[scale=0.35]{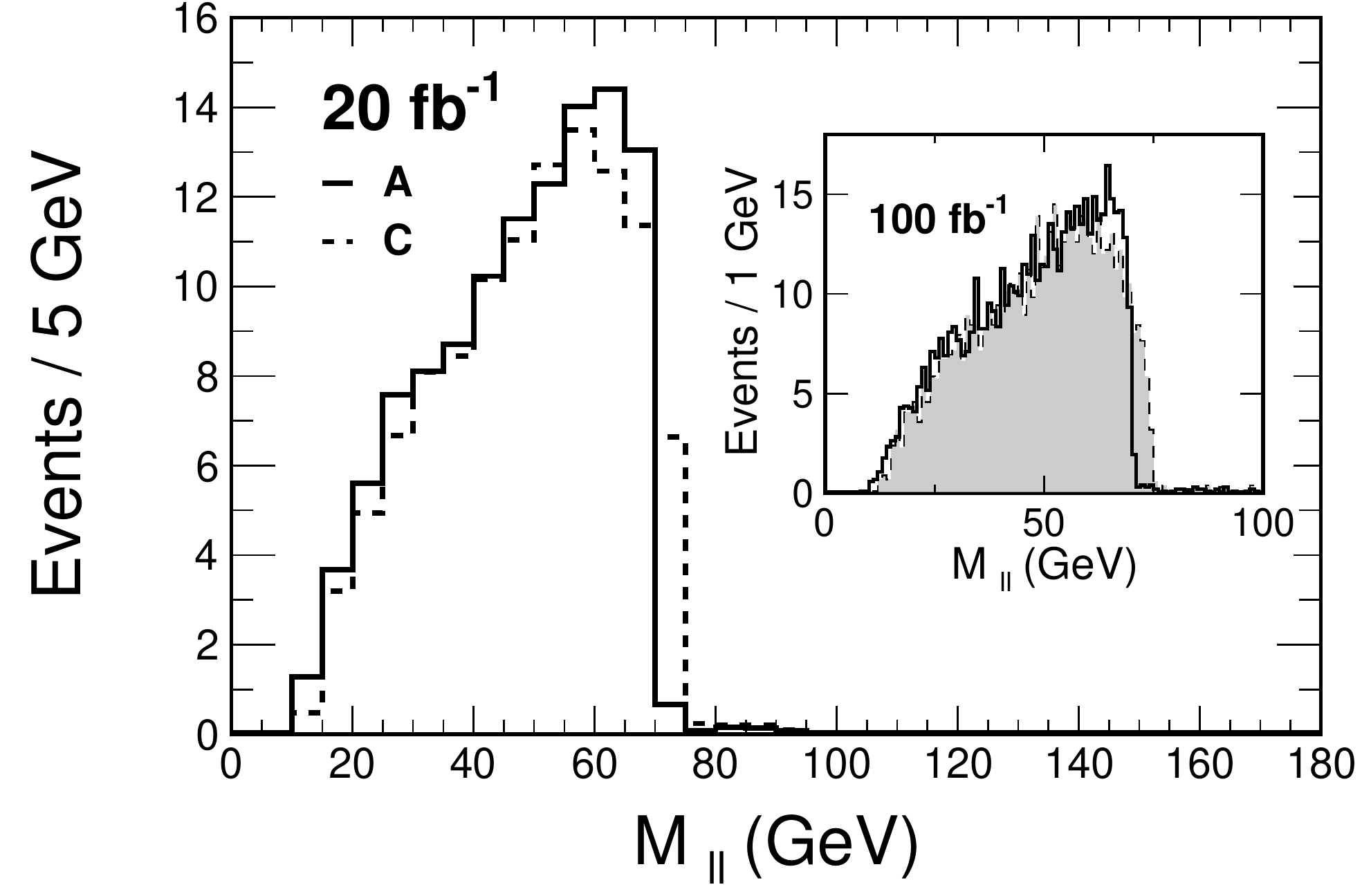}
\caption{Invariant mass distribution of the opposite sign same flavor  dileptons after implementing all selection cuts. 
The distributions are based on MC sample of 1000 fb$^{-1}$ integrated luminosity but normalized to 20 and 100 fb$^{-1}$ integrated luminosities.
}
\label{fig:DileptonEdgeCMS8}
\end{figure}

At all three benchmark points, a dilepton signal of around 120 events are obtained.  The dilepton mass distributions for points A and B are similar, and they both show a clear dilepton invariant mass endpoint at $M_{\tilde{\chi}^0_2}-M_{\tilde{\chi}^0_1}$ as each of two leptons has a fixed 35 GeV energy, where the invariant mass of the dilepton maximizes. The endpoint for point C is less pronounced due to the possible unequal energy between the two leptons in the three body decay of $\tilde{\chi}^0_2$. In Fig.~\ref{fig:DileptonEdgeCMS8}, we show the distributions for points A and C. The distribution for point B is identical to that of point A and is not shown. As the lepton energy is well measured at the LHC, it is possible to discriminate point C from other points by performing shape analysis with finer binning and higher integrated luminosity, as is shown in the inset. The statistical fluctuations in Fig.~\ref{fig:DileptonEdgeCMS8} corresponds to a MC sample equivalent to $1000$ fb$^{-1}$ integrated luminosity. All three benchmark points have less than 4\% multilepton final states compared to the dilepton final states as shown in Table~\ref{tab:multi}. We have chosen $p_T>10$ GeV for the third and fourth lepton in the multilepton final states.

\begin{table}[h]
\begin{tabular}{c|c|c|c}
\hline \hline
Number of multilepton events at 20 fb$^{-1}$
& A & B & C\\\hline
Three lepton & 2.14 & 2.10 & 4.46 \\\hline
Four leptons&1.10&0.7&1.52\\
\hline \hline
\end{tabular}
\caption{Number of $3l$ and $4ljj+\met$ events at benchmark points A, B and C.}
\label{tab:multi}
\end{table}

\section{Discussion and Conclusion}
\label{sect:discussions}
The dilepton excess, if confirmed at the ongoing run of the LHC, will  be able to probe SUSY models by investigating other signals associated with the models. In the case of the MSSM, the dilepton excess requires the existence of a Higgsino-type second-lightest neutralino and a chargino for masses around 300 GeV. 14 TeV LHC can produce these Higgsinos directly.  However, in such a scenario, the direct dark matter detection cross-section can potentially be a problem due to an appreciable mixture of Bino and Higgsino in the LSP which would push us to use the negative sign of $\mu$. This problem can be avoided in the NMSSM which also exhibits other distinguishable features which we discuss below.

For point A, the NLSP can be mostly wino. In this case,  it is possible to produce winos directly and the production cross-sections by Drell-Yan and vector boson fusion processes are larger by almost a factor of 4 compared to that in the Higgsino NLSP scenarios. The current bound($\sim 300$ GeV~\cite{atlas,cms}) still allows this point since the mass difference between the $\tilde\chi^{\pm}_1,\, \tilde\chi^0_2$ and $\tilde\chi^0_1$ is small and $\neu{2} \rightarrow \tilde{l} l$ decay branching is less than what is assumed by the CMS and ATLAS. The ongoing run will probe this point which does not suffer from the direct detection constraint. 

For point B, more neutralino states are involved to explain the excess. In this case the Higgsinos, which constitute mostly the $\tilde\chi^0_{3,4}$, are lighter than winos and they can be also be produced directly at the LHC. The  350 GeV $\tilde\chi^0_{3}$  ($m_{\tilde\chi^0_4}$ is 410 GeV) has a 75\% decay BF into $\tilde\chi^0_3\rightarrow Z\tilde\chi^0_1$ and most of the remaining BF is into $\tilde\tau \tau$. The $\tilde\chi^{\pm}_1$ is at 333 GeV and has appreciable a BF into $W+\tilde\chi^0_1$.  We have already shown  that the sbottom decays mostly into  $\tilde\chi^0_2$ and subsequently into $\tilde\chi^0_1$, whereas $\tilde\chi^0_{3,4}$ and $\tilde\chi^{\pm}_1$ decays predominantly to $\neu{1}$. The direct productions of $\tilde\chi^0_2$ and $\tilde\chi^0_1$ are small since they are mostly singlino and Bino respectively. Therefore, at least  four neutralinos and one chargino will  be potentially discovered at the ongoing run from two different production processes and can be used in a complimentary way to check whether this model point explanation for the excess is correct.  In addition, just like the Bino-Higgsino case in the MSSM, this scenario  suffers from the direct detection constraints. However, flipping the sign of the  gaugino and Higgsino mass parameters, can suppress the nucleon-LSP scattering.

For both points A, B and the MSSM, the mass-gap needed for these points to explain the excess is about 35 GeV. It can be difficult to probe the sleptons directly in the current runs of the LHC due to relatively small mass-gap between $\tilde{l}$ and the LSP. Even boosting the system with additional jets does not help, since such searches lose their efficacy for a mass splitting above $\sim 25$ GeV~\cite{dutta2}.

In contrast, at point C, the sleptons are relatively heavy in order not to overproduce event with more than two leptons.  If the sleptons are found to be a lot heavier than the NLSP, then the point C will be needed to explain the CMS excess. However, the direct slepton production can be probed  up to $\sim 700$ GeV~\cite{Ajaib,slepton} at 3000 fb$^{-1}$ of integrated luminosity. However, since $\tilde\chi^0_2$ and $\tilde\chi^0_1$ are mostly Bino and singlino respectively, the direct production cross-sections of them are very small.  Further, unlike points A, B and MSSM, this point does not have any light chargino associated with the lightest neutralinos which can be probed at the LHC. Therefore, if no other lighter neutralinos and charginos are found but the endpoint in the dilepton distribution still persists, the point C will provide the explanation. This point also does not suffer from the direct detection problem.  It is also possible to discriminate point C from other points by performing shape analysis.

In conclusion, if the CMS excess is proved to be correct at the ongoing run of the LHC, it will be possible to find an explanation in the context of NMSSM, which is not ruled out by the direct detection experiments and with unique features in the neutralino sector compared to the MSSM. It will be feasible to establish these NMSSM scenarios by investigating the direct productions of neutralinos, charginos and sleptons.

\medskip 

\section{Acknowledgements}

We thank Peisi Huang for cross-checking the lepton reconstruction efficiency in the MSSM. We also thank Kuver Sinha and Keith Ulmer for discussions. B.D., T.K. and T.G. are supported by DOE Grant DE-FG02-13ER42020. Y.G. thanks the Mitchell Institute for Fundamental Physics and Astronomy for support. T.K. is also supported in part by Qatar National Research Fund under project NPRP 5-464-1-080. N. K.'s contribution was made possible by the facilities of the Shared Hierarchical Academic Research Computing Network (SHARCNET: www. sharcnet.ca) and Compute/Calcul Canada.


\begin{thebibliography}{99}
\bibitem{bib:CMSlepEdge}
  V.~Khachatryan {\it et al.}  [CMS Collaboration],
  J. of High Energy Physics {\bf 04}, 124 (2015)
.
  
\bibitem{Huang:2014oza} 
  P.~Huang and C.~E.~M.~Wagner,
  Phys.\ Rev.\ D {\bf 91}, no. 1, 015014 (2015)

\bibitem{Grothaus:2015yha} 
  P.~Grothaus, S.~P.~Liew and K.~Sakurai,
  arXiv:1502.05712 [hep-ph].

\bibitem{Allanach:2015ria} 
  B.~Allanach, A.~Alves, F.~S.~Queiroz, K.~Sinha and A.~Strumia,
  arXiv:1501.03494 [hep-ph].

\bibitem{Dhuria:2015hta} 
  M.~Dhuria, C.~Hati, R.~Rangarajan and U.~Sarkar,
  Phys.\ Rev.\ D {\bf 91}, no. 5, 055010 (2015)
  [arXiv:1501.04815 [hep-ph]].

\bibitem{dutta}B.~Dutta, T.~Kamon, N.~Kolev, K.~Sinha, K.~Wang and S.~Wu,
  Phys.\ Rev.\ D {\bf 87}, no. 9, 095007 (2013)
  [arXiv:1302.3231 [hep-ph]].

\bibitem{bib:nmssm}
  J.~R.~Ellis, J.~F.~Gunion, H.~E.~Haber, L.~Roszkowski and F.~Zwirner,
  Phys.\ Rev.\ D {\bf 39} (1989) 844.
  M.~Drees,
  Int.\ J.\ Mod.\ Phys.\ A {\bf 4}, 3635 (1989).
  L.~Durand and J.~L.~Lopez,
  Phys.\ Lett.\ B {\bf 217}, 463 (1989).


\bibitem{bib:LHCHiggs} 
  CMS Collaboration,
  ``Observation of a new boson with mass near 125 GeV in pp collisions at sqrt(s) = 7 and 8 TeV,''  
 [arXiv:1303.4571 [hep-ex]].  

\bibitem{NMSSM:report} 
  U.~Ellwanger, C.~Hugonie and A.~M.~Teixeira,
  Phys.\ Rept.\  {\bf 496}, 1 (2010)
  [arXiv:0910.1785 [hep-ph]].



\bibitem{Dutta:2014hma} 
  B.~Dutta, Y.~Gao and B.~Shakya,
  Phys.\ Rev.\ D {\bf 91}, no. 3, 035016 (2015)
  [arXiv:1412.2774 [hep-ph]].


\bibitem{wh-ID}
  J.~Fan and M.~Reece,
  JHEP {\bf 1310}, 124 (2013)
  [arXiv:1307.4400 [hep-ph]].

\bibitem{Ajaib} 
  M.~A.~Ajaib, B.~Dutta, T.~Ghosh, I.~Gogoladze and Q.~Shafi,
  arXiv:1505.05896 [hep-ph].


\bibitem{higgsino-DD} 
  M.~Perelstein and B.~Shakya,
  JHEP {\bf 1110}, 142 (2011)
  [arXiv:1107.5048 [hep-ph]];
  M.~Perelstein and B.~Shakya,
  Phys.\ Rev.\ D {\bf 88}, no. 7, 075003 (2013)
  [arXiv:1208.0833 [hep-ph]].



 \bibitem{dutta3} 
  J.~L.~Feng and D.~Sanford,
  JCAP {\bf 1105}, 018 (2011).
R.~L.~Arnowitt, B.~Dutta and Y.~Santoso,
  Nucl.\ Phys.\ B {\bf 606}, 59 (2001).
J.~R.~Ellis, A.~Ferstl and K.~A.~Olive,
  Phys.\ Rev.\ D {\bf 63}, 065016 (2001)
  
\bibitem{bib:nmssmtools} 
  U.~Ellwanger, J.~F.~Gunion and C.~Hugonie,
 J. of High Energy Physics {\bf 02}, 066 (2005)
  U.~Ellwanger and C.~Hugonie,
  Comput.\ Phys.\ Commun.\  {\bf 175}, 290 (2006)


\bibitem{bib:Madgraph5}
J.~Alwall, M.~Herquet, F.~maltoni, O.~Mattelaer and T.~Stelzer,
``Madgraph 5: going beyond'',
J. of High Energy Physics {\bf 06}, 128 (2011);
  J.~Alwall, R.~Frederix, S.~Frixione, V.~Hirschi, F.~Maltoni, O.~Mattelaer, H.-S.~Shao and T.~Stelzer {\it et al.},
  J. of High Energy Physics {\bf 07}, 079 (2014)
  [arXiv:1405.0301 [hep-ph]].


\bibitem{CTEQ6.6} 
  P.~M.~Nadolsky, H.~L.~Lai, Q.~H.~Cao, J.~Huston, J.~Pumplin, D.~Stump, W.~K.~Tung and C.-P.~Yuan,
  Phys.\ Rev.\ D {\bf 78}, 013004 (2008)
  [arXiv:0802.0007 [hep-ph]].
 

\bibitem{bib:Pythia6}
T.~Sjostrand, S. Mrenna and P. Skands,
``Pythia 6.4 physics and manual'',
J. of High Energy Physics {\bf 05}, 026 (2006). 

\bibitem{bib:PGS4}
PGS4 is a parametrized detector simulation. We use release  120611 (\url{http://www.physics.ucdavis.edu/~conway/research/software/pgs/pgs4-general.htm}) in the CMS detector configuration.

\bibitem{MLM}
  M.~L.~Mangano, M.~Moretti, F.~Piccinini and M.~Treccani,
  J. High Energy Phys. 01 (2007) 013
  [hep-ph/0611129].



\bibitem{bib:ProspinoCode}
W.~Beenakker, R.~Hopker and M.~Spira, ``PROSPINO: A Program for the production of supersymmetric particles in next-to-leading order QCD'', arXiv:hep-ph/9611232 [hep-ph].

\bibitem{bib:ATLAS_Z}
  G.~Aad {\it et al.}  [ATLAS Collaboration],
  arXiv:1503.03290 [hep-ex].

\bibitem{cms}V.~Khachatryan {\it et al.}  [CMS Collaboration],
  Eur.\ Phys.\ J.\ C {\bf 74}, no. 9, 3036 (2014)
  [arXiv:1405.7570 [hep-ex]].



\bibitem{atlas}G.~Aad {\it et al.}  [ATLAS Collaboration],
  J. of High Energy Physics {\bf 04}, 169 (2014)
  [arXiv:1402.7029 [hep-ex]];
  G.~Aad {\it et al.}  [ATLAS Collaboration],
  J. of High Energy Physics {\bf 05}, 071 (2014)
  [arXiv:1403.5294 [hep-ex]].


\bibitem{dutta2} B.~Dutta, T.~Ghosh, A.~Gurrola, W.~Johns, T.~Kamon, P.~Sheldon, K.~Sinha and K.~Wang {\it et al.},
  Phys.\ Rev.\ D {\bf 91}, no. 5, 055025 (2015)
  [arXiv:1411.6043 [hep-ph]];
  Z.~Han and Y.~Liu,
  arXiv:1412.0618 [hep-ph];
  A.~Barr and J.~Scoville,
  J. of High Energy Physics {\bf 04}, 147 (2015) 
  [arXiv:1501.02511 [hep-ph]].


\bibitem{slepton} 
  J.~Eckel, M.~J.~Ramsey-Musolf, W.~Shepherd and S.~Su,
  J. of High Energy Physics {\bf 11}, 117 (2014) 
  [arXiv:1408.2841 [hep-ph], arXiv:1408.2841].




\end{thebibliography}
\end{document}